\documentclass[12pt,draftclsnofoot,onecolumn]{IEEEtran}
\usepackage{amssymb}
\usepackage{amscd}
\usepackage{amsmath}
\usepackage{epsfig}
\usepackage{amsthm}
\usepackage{graphics}
\usepackage{psfrag}
\usepackage{rotating}
\usepackage{amsmath} 
\usepackage{amsfonts}
\usepackage{url}
\usepackage{color}
\usepackage{epstopdf}
\usepackage{amsthm}
\usepackage{tikz}
\usepackage{mathtools}
\usepackage{multirow}
\usepackage[final]{pdfpages}
\usepackage{enumitem}
\usepackage{multicol}
\usepackage[nolist]{acronym}
\usepackage[font= small]{caption} 
\usepackage{subcaption}
\usepackage[ruled]{algorithm2e}
\usepackage{ragged2e}

\usepackage{algpseudocode}

\usepackage{pgfplots}
\usepackage{pgfplotstable}
\usepackage{cite}
\pgfplotsset{compat=1.12}
\usepgfplotslibrary{polar}
\usepackage{forest}

\newtheorem{remark}{Remark}

\DeclareMathAlphabet{\pazocal}{OMS}{zplm}{m}{n}

\newfont{\bbb}{msbm10 scaled 700}

\newfont{\bb}{msbm10 scaled 1100}
\newcommand{\CC}{\mbox{\bb C}}

\newcommand{\RR}{\mbox{\bb R}}

\newcommand{\Ss}{\mbox{\bb S}}

\newcommand{\av}{{\bf a}}

\newcommand{\fv}{{\bf f}}

\newcommand{\rv}{{\bf r}}
\newcommand{\sv}{{\bf s}}

\newcommand{\uv}{{\bf u}}
\newcommand{\wv}{{\bf w}}

\newcommand{\xv}{{\bf x}}
\newcommand{\yv}{{\bf y}}

\newcommand{\Gm}{{\bf G}}
\newcommand{\Hm}{{\bf H}}

\newcommand{\Um}{{\bf U}}

\newcommand{\Xm}{{\bf X}}

\newcommand{\Hc}{{\cal H}}

\newcommand{\thetav}{\hbox{\boldmath$\theta$}}

\newcommand{\Psim}{\hbox{\boldmath$\Psi$}}

\renewcommand{\Re}{{\rm Re}}

\newcommand{\herm}{{\sf H}}

\newcommand{\CB}{\mathcal{CB}_u}
\newcommand{\Ucs}{\Um_{\mathrm cs}} 

\newcommand{\Pav}{P_{\rm avg}}
\newcommand{\Na}{N_{\rm a}}
\newcommand{\Nrf}{N_{\rm rf}}
\newcommand{\Ntr}{N_{\rm tr}}

\usepackage{hyperref}
\hypersetup{
    bookmarks=true,         
    unicode=false,          
    pdftoolbar=true,        
    pdfmenubar=true,        
    pdffitwindow=false,     
    pdfstartview={FitH},    
    pdfnewwindow=true,      
    colorlinks=true,       
    linkcolor=red,          
    citecolor=cyan,        
    filecolor=blue,      
    urlcolor=blue           
}

\newcommand{\matr}[1]{{\mathbf #1}}

\usepackage{siunitx}
\DeclareSIUnit{\belmilliwatt}{Bm}
\DeclareSIUnit{\belsquaremeter}{Bsm}
\usepackage{mathtools}

\title{A Beam-Space Active Sensing Scheme for Integrated Communication and Sensing Applications}

\author{Saeid K. Dehkordi$^1$, Giuseppe Caire$^1$
\thanks{ $^1$ Communications and Information Theory Chair, Technical University of Berlin, Germany.}
\thanks{Corresponding author: s.khalilidehkordi@tu-berlin.de}
}

\begin{document}

\maketitle
\begin{acronym}
    \acro{ISAC}{Integrated Sensing and Communication}
	\acro{AWGN}{additive white Gaussian noise}
	\acro{HDA}{Hybrid Digital-Analog }
	\acro{BA}{beam acquisition}
	\acro{GEVT}{Generalized Extreme Value Theory}
	\acro{MIMO}{Multiple-input multiple-output}
	\acro{OTFS}{orthogonal time frequency space}
	\acro{SNR}{signal-to-noise ratio}
	\acro{mmWave}{millimeter wave}
	\acro{ML}{maximum likelihood}
	\acro{V2X}{vehicle-to-everything}
	\acro{OFDM}{orthogonal frequency division multiplexing}
	\acro{FMCW}{frequency modulated continuous wave}
	\acro{LoS}{line-of-sight}
	\acro{ISFFT}{inverse symplectic finite Fourier transform}
	\acro{SFFT}{symplectic finite Fourier transform}
	\acro{HPBW}{half-power beamwidth}
	\acro{ULA}{uniform linear array}
	\acro{CRLB}{Cram\'er-Rao Lower Bound}
	\acro{RF}{radio frequency}
	\acro{BF}{beamforming}
	\acro{RMSE}{root MSE}
	\acro{AoA}{angle of arrival}
	\acro{ISI}{inter-symbol interference}
	\acro{SI}{self-interference}
	\acro{TDD}{time division duplex}
	\acro{Tx}{transmitter}
	\acro{Rx}{receiver}
	\acro{SIC}{successive interference cancellation}
	\acro{PD}{probability of detection}
	\acro{HDA}{hybrid digital-analog}
	\acro{PSD}{power spectral density}
	\acro{FWHM}{full width at half maximum}
	\acro{SLL}{side lobe level}
	\acro{BS}{Base Station}
    \acro{FoV}{Field of View}
    \acro{CFAR}{Constant False Alarm Rate}
    \acro{OSCFAR}{Ordered Statistic Constant False Alarm Rate}
    \acro{UE}{User Equipment} 
\end{acronym}


\begin{abstract}
In this paper, we develop an active sensing strategy for a \ac{mmWave} band \ac{ISAC} system adopting a realistic \ac{HDA} architecture. To maintain a desired SNR level, initial \ac{BA} must be established prior to data transmission. In the considered setup, a \ac{BS} Tx transmits data via a digitally modulated waveform and a co-located radar receiver simultaneously performs radar estimation from the backscattered signal. In this \ac{BA} scheme a single common data stream is broadcast over a wide angular sector such that the radar receiver can detect the presence of not yet acquired users and perform coarse parameter estimation (angle of arrival, time of flight, and Doppler). As a result of the \ac{HDA} architecture, we consider the design of multi-block adaptive RF-domain ``reduction matrices'' (from antennas to RF chains) at the radar receiver, to achieve a compromise between the exploration capability in the angular domain and the directivity of the beamforming patterns. Our numerical results demonstrate that the proposed approach is able to reliably detect multiple targets while significantly reducing the initial acquisition time.  
\end{abstract}


\section{Introduction}\label{sec:Introduction}
 \ac{ISAC} applications have emerged as key enablers for 5G and beyond wireless systems to deal with challenging requirements in terms of spectral efficiency, localization, and power consumption among others\cite{ISAC_Survey}. In \ac{mmWave} communications, it is crucial to compensate the large isotropic path-loss  with highly directional \ac{BF} gain. This requires fast and accurate initial \ac{BA} to be established before data transmission (see e.g. \cite{song2018scalable} and references therein). In this work we focus on automotive applications where a \ac{Tx} unit, e.g. a \ac{BS} as a road-side infrastructure, communicates with other vehicles. In such applications, \ac{BA} for new users entering the \ac{FoV} is particularly challenging. Furthermore, \ac{BA} is a prerequisite for beam tracking and refinement \cite{Pedraza} of already acquired users such that the \ac{BS} can continually update the best beam for the users. 
 \begin{figure}
	\centering
	\includegraphics[width=3.5cm]{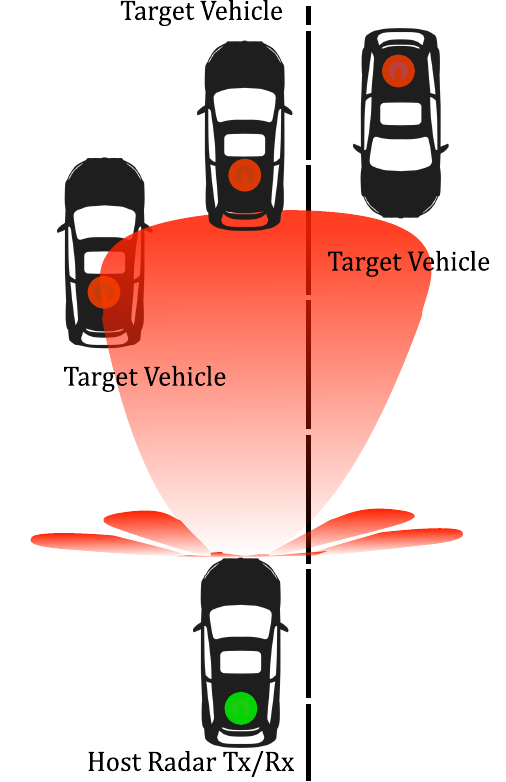}  
	\caption{ Discovery mode, where a Tx (a base station or a car) broadcasts a common message exploring a wide angular sector}
	\label{fig:Operational-Mode}
\end{figure}
 In our previous works \cite{gaudio2020effectiveness,Dehkordi_OTFS_ICC,OTFS_MIMO_AV_22}, we studied the joint target detection and parameter estimation problem with a \ac{BS} enhanced by a co-located Radar receiver, using \ac{OTFS}, i.e. a multi-carrier modulation proposed in \cite{hadani2017orthogonal} and applied to different \ac{MIMO} configurations (see, e.g., \cite{shen2019channel,ramachandran2018mimoOTFS}). As an extension to the aforementioned, here we consider improved initial target detection schemes. In this scheme, intended in the so called \textit{Discovery mode} presented in \cite{OTFS_MIMO_AV_22}, an \ac{OTFS} modulated signal is broadcast over a wide angular sector (fig.~\ref{fig:Operational-Mode}). The goal of the radar receiver is to detect the presence of targets (vehicles) that are not yet acquired, as well as estimating their relevant parameters (angle of arrival, range, and speed).\par
 The initial \ac{BA} for \ac{HDA} architectures has previously been studied in the context of a communication systems whereby the goal is to align a mobile \ac{UE} with a \ac{BS}. A major difference between the radar use case and the communication based case is the unavailability of direct feedback between the two entities, i.e., the radar receiver only relies on backscattered signals. The work of \cite{hiePM1} builds upon that of \cite{Alkhateeb_hie}, where hierarchical beamforming codebooks are used to narrow down the angular location of the \ac{UE}. The major issue for the bisection scheme in both works is that the sensing is initialized by sensing the wide \ac{FoV} with two (almost) constant gain beam patterns, which divide the \ac{FoV} in to equal sections. With the assumption of a constant available transmit power, this in effect leads to a very low \ac{BF} gain in the initial sensing stage, especially problematic for radar uses where signal power attenuates heavily with distance,i.e., $\propto 1/ r^4$. While \cite{hiePM1} improves the bisection scheme in \cite{Alkhateeb_hie} by applying a Posterior Matching scheme for the selection of codewords at each level, the assumption placed on known \ac{SNR} and channel coefficients is very impractical in the radar scenario. Additionally, the same work assumes a grided approach where more refined estimates, i.e., increased resolution leads to increasingly more levels of beampatterns thus increasing the acquisition induced latency. Perhaps the most significant drawback of these schemes for the radar use case, is the inability to simultaneously detect multiple users(targets). Furthermore, the work of \cite{hiePM1} has not extended the scheme for multiple RF chains in the \ac{HDA} architecture. The presented scheme in this work closely follows that of  resource management methods in cognitive radar, especially in the spatial domain, where the sensing output of the radar is used to improve the sensing pattern of the radar at the consequent probing periods\cite{Cognitive_Raei}. The main contributions of this work are summarized as follows. 

\begin{enumerate}
    \item We propose the use of multiple randomized reduction matrices across the processing interval, which are generated sequentially such that the matrix at block $b$ depends on the received signal in previous blocks. The proposed sensing strategy, whose role is to \textit{actively} sample the \ac{FoV} in the angular domain, leads to an improvement in detection probability. Additionally, this results in an improvement in parameter estimation performance by increasing the gain of the received signal across multiple blocks where targets are present.
    \item We propose a single level codebook as compared to the hierarchical codebooks of \cite{Alkhateeb_hie}. Besides reducing the system complexity, this design ensures a constant level of BF gain at each individual sensing level and improves the total gain via integration. Additionally, due to the ML estimation framework, we place no limitations on the resolution of the estimates, since as opposed to \cite{hiePM1}, the \ac{AoA} estimated is not determined by the codebook refinement level.
    \item The proposed scheme is easily adjustable with the number of available RF chains in a given \ac{HDA} system.
\end{enumerate}



\section{System model}\label{sec:phy-model}

\subsection{Physical Model}\label{subsection:PhysicalModel}
We consider a system operating over a channel bandwidth $W$ at the carrier frequency $f_c$. A \ac{BS} Tx is equipped with an \ac{ULA} of $\Na$ elements with  $\Nrf$ Tx RF chains ($\Na \gg \Nrf$), and a radar receiver co-located with the \ac{BS}. For simplicity of exposition, we assume that the Tx array and the Rx radar array coincide and that the Tx and Rx signals are separated by means of full-duplex processing.\footnote{Full-duplex operations can be achieved with sufficient isolation between the transmitter and the (radar) detector and possibly interference analog pre-cancellation in order to prevent the (radar) detector saturation \cite{sabharwal2014band}.} 
We consider a point target model, such that each target can be represented by a \ac{LoS} path only \cite{kumari2018ieee,nguyen2017delay,grossi2018opportunisticRadar}.
This model can be justified for \ac{mmWave} channels as they incur large isotropic attenuation such that 
all multipath components between the \ac{BS} and each target receiver disappear below the noise floor after reflection. 
By letting $\phi\in [-\frac{\pi}{2}, \frac{\pi}{2}]$ be the steering angle and 
considering a \ac{ULA} with $\lambda/2$ spacing, the Tx/Rx array response are given by: 
\begin{align} \label{eq: ULA_Mnf}
	&[\av(\phi)]_{i} = e^{j\pi(i-1)\sin(\phi)},\quad i\in1,\dots,\Na
\end{align}
Since this paper focuses on the radar processing, we consider the channel model for the backscattered signal. The channel for the backscattered signal with $P$ targets is given by the superposition of $P$ rank-1 channel matrices, each of which corresponds to the \ac{LoS} propagation from the Tx array to each target and back to the radar Rx array along the same \ac{LoS} path. This results in the $\Na\times\Na$ time-varying MIMO channel given by \cite{vitetta2013wireless}
\begin{align}\label{eq:Channel}
	\Hm (t, \tau) =  \sum_{p=0}^{P-1} h_p \av(\phi_p) \av^\herm (\phi_p)\delta(\tau-\tau_p)  e^{j2\pi \nu_p t}\,,
\end{align}
where 
$h_p$ is a complex channel gain including the \ac{LoS} pathloss and the radar cross-section coefficient\cite{richards2014fundamentals}:
\begin{align} \label{eq:Chanel_coeff}
	|h_{p}|^{2} = \frac{\lambda^{2}\sigma_{{\rm rcs}, p}}{(4\pi)^{3}d^{4}_{p}}~,
\end{align}
where $\lambda=\frac{c}{f_c}$ is the wavelength, $c$ is the speed of light, $\nu_p$ is the round-trip Doppler shift,
$\tau_p$ is the round-trip delay (time of flight for a distance of $d_{p}$), $\phi_p$ denotes the \ac{AoA}, and $\sigma_{{\rm rcs}, p}$ is the radar cross section (RCS) in $\mathrm{m}^2$, corresponding to the $p$-th target.  We assume that the channel parameters $\{h_p, \phi_p, \nu_p, \tau_p\}_{p=1}^P$ remain constant over the coherence processing interval $T_{\mathrm CPI}$ of $B$ time-frequency blocks, where each time-frequency block is the product of a total bandwidth $W$ [Hz] and a block duration of $T_{\rm block}$.

\subsection{Digital Modulation Scheme: \ac{OTFS}}\label{subsec:OTFS-Input-Output}
The presented beamforming framework in this paper is not specific to a digital modulation scheme and in principle could be used in any digital modulation (e.g. OFDM, \ac{OTFS}, etc.) as long as it is compatible with a \ac{HDA} architecture. In line with our previous works, we consider \ac{OTFS} in the following. The \ac{OTFS} modulation format, offers robustness to high Doppler shifts and operates efficiently in the presence of Doppler-delay-domain sparse channels \cite{gaudio2020effectiveness}. In \ac{OTFS}, the total bandwidth is divided into $M$ subcarriers with separation $\Delta f$ such that $W = M\Delta f$. $T$ denotes the symbol time and $N$ is the number of \ac{OTFS} symbols per block, yielding a block duration of $T_{\rm block}=NT$. We also consider $T\Delta f = 1$, which is typical in most \ac{OTFS} literature \cite{gaudio2020effectiveness,hadani2017orthogonal,raviteja2018interference}. 
The data symbols belonging to some QAM constellation are given as $\{\xv_{k,l} \in \CC^{} : k=0,\dots,N-1, \; l=0,\dots,M-1\}$. The Tx applies the \ac{ISFFT}, converting the Doppler-delay domain data block $\{\xv_{k,l}\}$ into the corresponding time-frequency data block $\{\Xm[n, m]\}$, and the $\Nrf$-dimensional continuous-time received signal at the radar Rx is given by 
\begin{align}\label{eq:Received-Signal-First}
	\rv(t) = \sum_{p=0}^{P-1} h_p {\Um}^\herm \av(\phi_p) \av^\herm(\phi_p) \fv s(t - \tau_p) e^{j2\pi \nu_pt}\,.
\end{align}
where $\sv(t)$ is the continuous-time transmit signal obtained by a linear mapping from \{\Xm[n, m]\}. Due to space limitation, readers are referred to our previous work \cite{gaudio2020effectiveness} for the detailed derivation. The design of the Rx \ac{BF} matrices, denoted $\Um\in \CC^{\Na \times \Nrf}$ is the topic of interest in this work. During \ac{BA}, the \ac{BS} sends a single data stream through a \ac{BF} vector $\fv$, designed to uniformly cover a given (wide) angular sector as the \ac{BS} has no a priori knowledge of the location of the targets. At a given Doppler-delay pair $(k,l)$, the channel output without noise is 
\begin{align}\label{eq:381}
    \yv[k,l] &= \sum_{p=0}^{P-1} h_p' {\Um}^\herm\av(\phi_p)\av^\herm(\phi_p)\fv \sum_{k', l'} x_{k',l'} \Psi_{k, k',l,l'}(\nu_p,\tau_p) ,
\end{align}
where $h'_p=h_p e^{j2\pi \tau_p \nu_p}$. $\Psi_{k, k',l,l'}(\nu_p,\tau_p) $, i.e. the Doppler-delay
crosstalk coefficient, denotes the $p$-th target OTFS modulated channel response 
at Doppler-delay index $[k,l]$ relative to a symbol at $[k',l']$
(see \cite{gaudio2020effectiveness}).
In this work, $\fv$ is designed by a semidefinite relaxation of the magnitude least-squares  problem described in \cite{OTFS_MIMO_AV_22}, allowing a desired beamformer $\fv$ with very minimal ripples within the main beam. The model in \eqref{eq:381} corresponds to a single-user MIMO channel.
\subsection{Detection}\label{sec:Detection and Estimation} 
For radar detection, we consider multi-block processing across $B$ blocks, where the input-output relation in \eqref{eq:381} holds for each block. While the Tx beamforming matrix $\fv$, illuminating a wide \ac{FoV}, remains constant over the blocks, the reduction matrix at block $b$ denoted $\Um_b,~  b \in [B]$, varies from one block to another, such that the design of $\Um_b$ is influenced by the received signal at the Radar Rx in the previous blocks ($1,...,b-1$). The considered framework consists of two detection stages, namely a single block detection where presence of targets is evaluated at each individual block and a final detection performed over the integrated signal from all $B$ blocks. Let $\mathring{\thetav} = \{\mathring{h}_{p}, \mathring{\nu}_{p}, \mathring{\tau}_{p}, \mathring{\phi}_{p}\}$ and 
$\thetav = \{h_p, \nu_p, \tau_p, \phi_p \}$ denote the true and the hypothesized parameter values of the targets, respectively. The received signal expression (\ref{eq:381}) can be written in a compact form by blocking the $NM$ Doppler-delay signal components into an $NM \times 1$ vector. To avoid notation ambiguity, we use underline to denote blocked quantities. For each $b = 1, \ldots, B$, the  effective channel matrix of dimension ${\Nrf} NM\times NM$ is defined 
\begin{align}\label{eq:G-Matrix}
	\underline{\Gm}_{b}(\nu, \tau,\phi)\triangleq\left({\Um}^H_b\av\left(\phi\right)\av^H(\phi)\fv \right)\otimes\Psim (\nu,\tau)\,,
\end{align} 
where $\Psim(\nu,\tau)$ is defined such that $[\Psim(\nu,\tau)]_{kM+l, k’M+l'}= \Psi_{k, k’,l, l’}(\nu,\tau)$ for $k, k’\in [0, N-1]$ and $l,l'\in [0, M-1]$, 
where $\Psi_{k,k',l,l'}(\nu,\tau)$ is defined in \cite{OTFS_MIMO_AV_22}, and
$\otimes$ is the Kronecker product. By stacking the $N\times M$ \ac{OTFS} symbol block 
into a $NM$-dimensional vector $\underline{\xv}_b$ and defining the blocked 
output vector $\underline{\yv}_b$ of dimension ${\Nrf} NM \times 1$, the received signal takes on the form 
\begin{align}\label{eq:Received-signal}
\underline{\yv}_b = \left ( \sum_{p=0}^{P-1} \mathring{h}_{p} \underline{\Gm}_{b}(\mathring{\tau}_{p}, \mathring{\nu}_{p},\mathring{\phi}_{p}) \right ) \underline{\xv}_b  + \underline{\wv}_b, \;\;\; b=1,\dots, B,
\end{align}
where $\underline{\wv}_b$ denotes the \ac{AWGN} vector with independent and identically distributed entries of zero mean and variance $\sigma_w^2$. 
Note that the number of targets $P$ is unknown and they are simultaneously illuminated by a single wide \ac{FoV} beacon signal. The target detection problem can be formulated as a standard Neyman-Pearson hypothesis testing problem \cite{VPoor} for which the solution that maximizes the detection probability subject to a bound on the false-alarm probability is given by the Likelihood Ratio Test
\begin{equation}
    \ell(h_p,\nu_p,\tau_p,\phi_p)  \underset{\Hc_F}{\overset{\Hc_T}{\gtrless}} T,
\end{equation}
where the threshold $T$ determines the tradeoff between detection and false-alarm probabilities. $\Hc_T$ ans $\Hc_F$ are generic \textit{True} and {False} hypotheses. In this work we use $\Hc_p$/$\Hc_n, T_b$ and $\Hc_1$/$\Hc_0, T_r$ pairs to distinguish between the different hypotheses and thresholds in the single block and $B$-block testing procedures, respectively.  
Since the true value of the parameters is unknown, we use the Generalized Likelihood Ratio Test
\begin{equation}
    \max_{h_p,\nu_p,\tau_p,\phi_p} \; \ell(h_p,\nu_p,\tau_p,\phi_p)  \underset{\Hc_F}{\overset{\Hc_T}{\gtrless}} T.
\end{equation}

\noindent Neglecting the arguments in $\underline{\Gm}_{b}(\tau_p, \nu_p,\phi_p)$ to avoid excessive clutter in the notation, the log-likelihood ratio (LLR) for the binary hypothesis testing problem, multiplied by $\sigma_w^2$ for convenience, is given by 
\begin{align}
    \ell(& h_p,\nu_p,\tau_p,\phi_p)  = \sigma_w^2 \log \frac{\exp\left ( - \frac{1}{\sigma_w^2} \sum_{b=1}^B \left \| \underline{\yv}_b - h_p \underline{\Gm}_b \underline{\xv}_b \right \|^2 \right )}{\exp \left ( - \frac{1}{\sigma_w^2} \sum_{b=1}^B \| \underline{\yv}_b \|^2 \right )} \nonumber \\
    & = 2 \Re\left \{ \left ( \sum_{b=1}^B \underline{\yv}_b^\herm \underline{\Gm}_b \underline{\xv}_b \right ) h_p \right \} - |h_p|^2 \sum_{b=1}^B \| \underline{\Gm}_b \underline{\xv}_b \|^2 \label{log-likelihood2}
\end{align}

\noindent The maximization of \eqref{log-likelihood2} with respect to $h_p$ for fixed $\tau_p, \nu_p, \phi_p$ is immediately obtained  as
\begin{equation} 
\widehat{h}_p = \frac{\left ( \sum_{b=1}^B \underline{\yv}_b^\herm \underline{\Gm}_b \underline{\xv}_b \right )^*}{\sum_{b=1}^B \|\underline{\Gm}_b \underline{\xv}_b\|^2}.
\label{opth}
\end{equation}
Replacing \eqref{opth} into \eqref{log-likelihood2} we obtain the LLR in the form:
\begin{equation}
    \ell(\widehat{h}_p,\nu_p,\tau_p,\phi_p) = \frac{\left | \sum_{b=1}^B \underline{\yv}_b^\herm \underline{\Gm}_b \underline{\xv}_b \right |^2}{\sum_{b=1}^B \| \underline{\Gm}_b \underline{\xv}_b \|^2}. \label{log-likelihood2S}
\end{equation}
where eq. \eqref{log-likelihood2S} is evaluated on a 3-dimensional discrete grid. Hereinafter, we define the function $S(\nu,\tau,\phi)$ given by \eqref{log-likelihood2S}
after replacing $\nu_p \leftarrow \nu,\tau_p \leftarrow \tau,\phi_p \leftarrow \phi$. Additionally, for notation brevity we use $T_r \leftarrow T_r(\nu,\tau,\phi)$ and $T_b \leftarrow T_b(\nu,\tau,\phi)$.  

\section{Active Sensing Strategy}\label{sec:ActiveBeamDesign}
  In this section, we discuss the design of the sequence of reduction matrices 
$\{\Um_b : b \in [1,\ldots,B]\}$. 
The goal is to achieve a good trade-off between exploration of the beam space and the directivity of the beam pattern in order to achieve good \ac{BF} gain in the explored directions. Given that no a priori information on the angular location of the targets is available during \ac{BA}, the coverage of a wide \ac{FoV} requires a very dense codebook of Fourier-type beamformers. Namely, for antenna arrays in the \ac{mmWave} regime with typically large $\Na$, only $\Nrf$ ( $\Nrf\ll\Na$) Fourier directions ($\Nrf$ dimensional projection of the beam-space) can be explored at each block, leading to a large number of blocks $B$ to cover the entire \ac{FoV} and therefore incurring a large latency for target detection. 
Hence, we consider a set of \textit{flat-top} beams designed to provide sufficiently large \ac{BF} gain (and therefore maintain a good operating SNR) over an extended angular span as compared to Fourier beams. Note that in comparison to the hierarchical method in \cite{hiePM1,Alkhateeb_hie}, the codewords in the presented scheme maintain a certain \ac{BF} gain. Let $\Omega$ denote the \ac{FoV} and $\mathcal{CB}_u \coloneqq (\uv_{1},...,\uv_{Q}) \in \CC^{\Na\times Q}$ a codebook with a set of $Q>\Nrf$ approximately orthogonal (i.e.$, ~ \uv_{q}^{H}  \uv_{p}\approx 0 ~ \forall q\neq p $) uniformly spaced flat-top beams each of width $\Delta = \Omega/Q$  (see for example Fig.~\ref{fig: active_strategy}). The choice of $\Delta$ is a design parameter striking a trade-off between BF gain and coverage. The reduction matrices $\Um_b = (\uv_{1,_{b}},...,\uv_{\Nrf,_{b}}) \in \CC^{\Na\times \Nrf}$, are constructed such that at the first block $b=1$, $\Nrf$ codewords are drawn at random from $\mathcal{CB}_u$ to \textit{partially} cover the beamspace $\Omega$ illuminated by $\fv$. Such a reduction matrix design can suffer two main drawbacks which we aim to overcome. First, if the angular \ac{FoV} is only partially covered, it is probable that some targets in unexplored sections will be missed. Second, as a direct consequence of the flat radiation pattern response of the beam, the ML estimation algorithm described in section \ref{sec:Detection and Estimation} will suffer in estimation accuracy. The proposed strategies in this section deal with these two effects without introducing increasing amounts of codebook complexity.

\subsection*{Strategy 1: Grid-Shifted (GS) beams} 
\noindent One of the issues arising with using multi-directional \ac{BF} patterns at the receiver, is that the sidelobes of the beams can add up constructively and lead to leakages between adjacent grid points. This means that a stronger target in one bin can leak into other grid points. This effect becomes more critical in the lower \ac{SNR} regimes or when a very strong target is present. To circumvent this effect we define a \textit{weighted} detection function $\Tilde{S}(.)$:

\begin{align}\label{block-level-prob_BP}
\forall \nu,\tau :~~ \Tilde{S}_{b}(\nu,\tau,\phi) & \coloneqq S_{b}(\nu,\tau,\phi)W_{\rm b}(\phi).
   \\ S_{b}(\nu,\tau,\phi) & = \frac{ |  \underline{\yv}_b^\herm \underline{\Gm}_b \underline{\xv}_b |^2}{ \| \underline{\Gm}_b \underline{\xv}_b \|^2}, \nonumber
\end{align}

where the weight function $W_{\rm b}(\phi) \in \RR^{1\times G}$ models the normalized beam pattern resulting from $\Um_b$ over the angular grid with cardinality $G$ (i.e., $ \phi_g \in [\phi_1,...,\phi_G]$), where we use the shorthand notation $ W_{\rm b}(\phi_g) = W(\phi=\phi_g|\Um_b)$.
 This ensures large signal values (regardless of position on the delay/Doppler grid) corresponding to angles where the intended receive beamforming power is weak, are attenuated. As an example, the dashed orange shapes in fig.~\ref{fig: plot_multi_block_Strg1}, show portions of the received signal that fall above the threshold, as a result of the effects explained above, but are not in the receive beamforming of the corresponding block.
 \par
\begin{figure}[t!]
	\centering
	\includegraphics[scale=0.7]{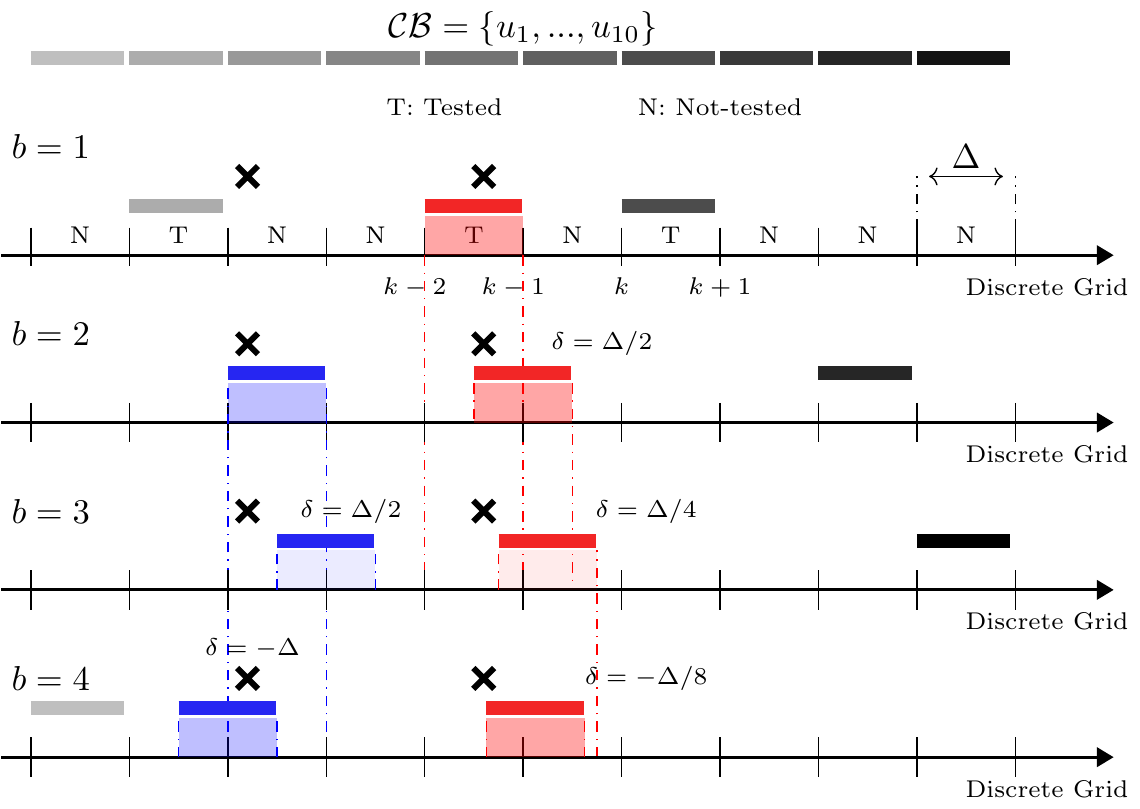}
	\caption{Strategy for 4 consecutive blocks with a Q=10 element codebook, $\Nrf = 3$ and two targets. The crosses indicate target positions.}
	\label{fig: active_strategy}
\end{figure}

 We resort to an exemplary scenario such as depicted in fig.~\ref{fig: active_strategy} to explain the GS scheme. Starting from the first block $b=1$, a \textit{hard-thresholding} based detection (refer to \ref{sec: hard_thr}) is applied to the ML metric \eqref{log-likelihood2S} resulting from $\Um_{\rm b=1}$:  
\begin{align} \label{block-level-hypo}
    \Tilde{S}_{b}(\nu,\tau,\phi)  \underset{\Hc_{n}}{\overset{\Hc_{p}}{\gtrless}} T_{b}
\end{align}
where $\Hc_{p}$ and $\Hc_{n}$ denote the positive and negative hypothesis, respectively. After evaluating \eqref{block-level-hypo}, the detected points are binned into discrete sections corresponding to the angular span of the codewords. Let $\Ntr \leq \Nrf$ be the number of detected targets. Here we define the subscript $l \in [L_{\mathrm max}]$, which defines the level-indexing for beams that should remain active. Define the $\Ntr$ beams wherein a target is initially ($l=0$) detected  as $\uv^{\star}_{0,j},~\uv^{\star} \subset \CB,~j \in \{1,...,\Ntr\}$.
In the next sensing block (in this example $b=2$), $\uv^{\star}_{0,j}$ are reserved and their pointing direction is partially shifted on the discrete grid to the right (or left) by a shift value $\delta_1( = \Delta/2)$, i.e. $\uv^{\star}_{1,j}(\phi) = \uv^{\star}_{0,j}(\phi-\delta_1)$. Intuitively speaking, in analogy to the hierarchical refinement of the codeword width in \cite{Alkhateeb_hie}, here, half of the beam (i.e. $\Delta/2$) is tested to further isolate the target. Additionally $\Nrf-\Ntr$ new codewords randomly sampled from the beamspace positions previously not covered (i.e. not tested) are selected and assigned to the available RF chains. These beams will constitute $\Um_{\rm b=2}$. Next, the sensing measurement is made and \eqref{block-level-hypo} is evaluated. Based on the decided hypothesis and setting $\delta_0=0,~\delta_1=\Delta/2$, the beam shifting values at the $l^{th},\; l> 2$ level are determined according to:
\begin{align}
 \delta_l & = \gamma_{l}(\delta_{l-1}) \nonumber \\
 \uv^{\star}_{l,j}(\phi) = \uv^{\star}_{l-1,j}&(\phi-\delta_l) 
  ,~~ \gamma_l = \begin{cases}
    1/2 & \;\; \small \text{if $\Hc_{p}$}\\
    -1/2 & \;\; \small \text{if $\Hc_{n}$} \,.
    \end{cases}
    \label{eq: strategy_lth}
\end{align}
For the case of $l=2$, we have $\Hc_{p} \mapsto \gamma_2= 1/2$ and $\Hc_{n} \mapsto \gamma_2= -2$. Eq.~\eqref{block-level-prob_BP} is repetitively evaluated for new detections at each block (see Fig.~\ref{fig: active_strategy}). The center angle shifting of the codeword is easily obtained by a phase rotation of the codeword (i.e. $e^{(-j\pi \cos{(\delta)}i)}\uv$ for the \ac{ULA} in \eqref{eq: ULA_Mnf} for a shift of $\delta$ in the beamspace). $L_{\mathrm max}$ and $B$ are design parameters. Our simulations show that $L_{\mathrm max} = 3$ or $4$ already leads to significant improvement in the peak shape, namely a peak sharpening effect occurs, and $B$ is adaptively selected until each section of the beamspace $\Omega$ has cumulatively been covered. Due to the assumption of invariant target parameters across time over $B$ blocks, the Rx receives a statistically equivalent measurement of the channel by using partially covered sections of $\Omega$ which are designed to cover it in $B$ frames. Finally, the received signal from all blocks is processed via \eqref{log-likelihood2S} and an OS-CFAR threshold $T_r$ (see \ref{sec: hard_thr}) to extract targets (see fig.~\ref{fig: plot_multi_block_Strg1}). 

\subsection*{Strategy 2: Circulant-Shifted (CS) beams}
\begin{figure}[h]
\centering
\includegraphics[scale=0.32]{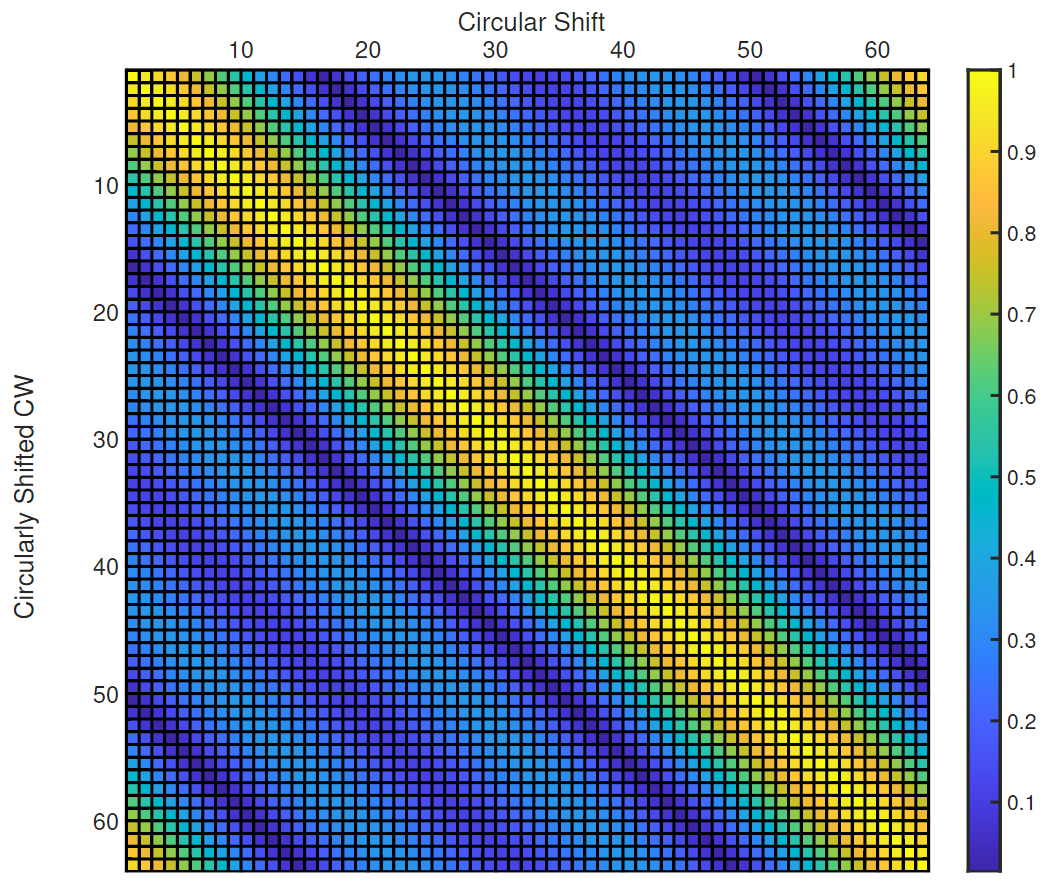}
\caption{Mutual Coherence function of an exemplary codeword (flat-top-beam) with $\Na=64$ elements from the used code book $\CB$. The $K^{th}$ row of this 'matrix' corresponds to the mutual coherence value of the $K$-positions circularly-shifted codeword to its shifted versions with circular shifts $\in [1,...,64]$.}
\label{fig: Mut_Coh}
\end{figure}

\noindent It is well known that applying a circular shift on the coefficients of a beamforming vector does not alter the magnitude of its radiation pattern in the angle domain \cite{Kumari_WideBeam}. To take advantage of this property for the active sensing scheme, for each codeword $\uv^{\star}$($\in \CB$) in which a target is initially detected, instead of using $L_{\rm max}$ grid-shifted beams, $L_{\rm max}$ circularly shifted variants of $\uv^{\star}$ are used in the following blocks. To this end, we are interested in constructing a dictionary $\Um^{\star}$ containing $L_{\rm max}$ highly incoherent atoms. Then, each of the $L_{\rm max}$ atoms are selected as a replacement for $\uv^{\star}$ in the proceeding blocks. The circulant-shifted codewords provide vector observations that allow an improved ML estimation. The random selection of codewords for the other RF chains, the weighting function and thresholding follow the same procedure as GS. Define $\Gamma(\uv^{\star},k)$ as an operator function inducing a circular shift of $k$ positions to $\uv^{\star}$. Then $\Ucs \in \CC^{\Na\times \Na}$ is a matrix containing $\Na-1$ shifted versions of the selected codeword $\uv^{\star}$,
\begin{equation}
  \Ucs=\left[\Gamma(\uv^{\star},0),\Gamma(\uv^{\star},1),..., \Gamma(\uv^{\star},\Na-1)\right].\\ \nonumber
\end{equation}
To construct the dictionary $\Um^{\star}$ , the Cumulative Coherence Function (also referred \textit{Babel} function)  \cite{Tropp2, Babel} which is a generalization of mutual coherence, and defined below is used.
\begin{equation} \label{CCF}
\mu(L)= \underset{|\Um^{\star}|=L}{\text{max}}~ ~\underset{i\in \Ucs \setminus \Um^{\star}}{\text{max}}\sum_{j\in \Um^{\star}} \vert \matr{u^{\star}_i}^H\matr{u^{\star}_j}\vert,
\end{equation}
For the problem in this work, CCF can be simplified to finding the $L = L_{\rm max}$ shifted versions with the lowest vector-wise mutual coherence, ( i.e. most orthogonal) corresponding to: 
\begin{align}\label{comb_prob}
\underset{}{\text{minimize}} &\sum_{i,j,~ i\neq j} \vert \Gamma(\uv^{\star},i)^H \Gamma(\uv^{\star},j) \vert \nonumber \\
s.t.~~ &\left| \Um^{\star}\right| = L_{\rm max}, ~~ i,j\in [0:\Na-1]
\end{align} 
The mutual coherence of an exemplary codeword of length $\Na=64$, is shown in fig.~\ref{fig: Mut_Coh}. From fig.~\ref{fig: Mut_Coh}, it can observed that the mutual coherence function is symmetric along both diagonals and therefore, the combinatorial-search problem in \eqref{comb_prob} can be significantly reduced.
For our particular problem with the simplifications mentioned above, we resort to a reduced brute-force search to solve this problem. For each codeword in $\CB$, the best $L = L_{\rm max}$ shifted versions are calculated and stored in a Look-Up-Table. Note that, works such as \cite{Babel} have proposed lower complexity solutions. Also noteworthy is that for larger problem sizes (e.g. larger $\Na$), it is possible to solve this problem using heuristics-based approaches such as Ant Colony optimization or the Genetic algorithm. 

\subsection{Hard-Thresholding}\label{sec: hard_thr}
 Two distinct thresholding techniques have been considered in this work. The adaptive threshold $T_r$ used for the multi-block ($B$ blocks) integrated signal, follows the OS-CFAR framework, and is obtained as in \cite{OTFS_MIMO_AV_22}. Due to the very low \ac{SNR} prior to block integration, we consider a \textit{hard-threshold} in the single block scenario. The detection threshold required to distinguish the signal component from noise has no closed-from solution, therefore by fixing a false alarm probability of $P_{\mathrm fa}= 10^{-3}$ in simulations and applying a technique based on the \ac{GEVT} \cite{Broadwater,Ozturk}, it is possible to obtain the hard threshold $T_b$ for single block detection. The \ac{GEVT} \ac{CFAR} detector is based on the fact that for a distribution of sufficiently large sample size that satisfies the Fisher-Tippet theorem, the conditional distribution of values over a given sufficiently high threshold, i.e. the \textit{tail} distribution, converges to the generalized Pareto distribution (GPD) (See \cite{Broadwater} and references therein for more details). The GPD is parameterized by a shape parameter which is an important factor in characterizing the nature of the distribution. Typically, the distributions considered in radar detection tend to be unbounded from the right. This implies a strictly non-negative shape parameter, corresponding to a Gumbel ($=0$) or Frechet ($>0$) distribution \cite{distbs}. In our problem, this parameter is not known and calculated from the simulated data. The algorithmic procedure to obtain the threshold $T_b$ in section \ref{sec:Detection and Estimation} is shown Alg.~\ref{alg: GEVT_CFAR}.  

\SetKwFor{For}{\hspace{-0.1cm}For}{do}{\hspace{-0.2cm}End}
\setlength{\algomargin}{0.3cm}
\begin{algorithm}
      \begin{minipage}{0.65\linewidth}
        \centering
	\SetAlgoLined
	\justifying
	\noindent
	\KwResult{Detection threshold ($T_b$).}
	\noindent\textbf{Inputs:} signal points $\Ss = \{s_0,...,s_N\}$ (detections), signal clipping threshold $\eta$ \;
	
	\For{Iteration $b=1,2,\dots,B$}{
		{\noindent\textbf{1) Fit GPD:}
		\begin{itemize}
		    \item 	Given $\eta$, remove all signal points $\leq \eta$
		    \item 	calculate the shape and scale parameters of the GPD:
		    \begin{align}\label{L-moments}
            \alpha = 2-\frac{d_0}{2d_1-d_0}\nonumber \\
            \beta = (1-\alpha)d_0
			\end{align}
			where $d_1,...,d_K$ are the detections $> \eta$, \\ $d_0 = 1/K \sum_{k=1}^{K}s_k$ and $d_1 = 1/K \sum_{k=1}^{K}\frac{k-1}{K-1}s_k$.
			\item Use GPD to approximate tail distribution
		\item Calculate tail probability $P_{\eta} = P(\Ss>\eta) = K/N$,\\ where N is the total number of detections.
		\item Define $P_{T} = P_{\mathrm fa}/P_{\eta}$ where $P_{T}$ is the probability that the detection are larger than a desired threshold $T$.
		
			\begin{align}\label{eq: tail_threshold}
              T = s_{N-K+1}+ \frac{\beta}{\alpha}\left( (1-\frac{N}{K})^{-\alpha}-1\right)
			\end{align}
		
		\end{itemize}

		}
		\noindent\textbf{2) set $T_b=T$}\\
	}
	\caption{\textit{Hard-Threshold calculation via \ac{GEVT}}}
	\label{alg: GEVT_CFAR}
	\end{minipage}
\end{algorithm}


\section{Numerical Results}\label{sec:Numerical-Results}
\renewcommand{\arraystretch}{1.3}

\begin{table}
	\caption{System parameters}
	\centering
	\scalebox{0.99}{
	\begin{tabular}{|c|c|}
		\hline
		$N=64$ & $M=64$ \\ \hline
		$f_c=30.0$ [GHz] & $W=150$ [MHz] \\ \hline
		$\Pav=24$ [dBm] & $\sigma_{\mathrm{rcs}}=1$ [m$^2$] \\ \hline
		Noise Figure (NF) $=3$ [dB] & Noise PSD $(\eta_{\text{PSD}})$ = $2\cdot10^{-21}$ [W/Hz] \\ \hline
		$\Na=64$ & $\Nrf=4$ \\ \hline
	\end{tabular}
	}
	\label{tab:System-Parameters}
\end{table}

\subsection{Simulation Setup}
The system parameters for the simulations in this section are provided in Table \ref{tab:System-Parameters}. We assume a single \ac{LoS} path between the Tx and the radar target. The radar two-way pathloss is defined as \cite[Chapter 2]{richards2014fundamentals} $\mathrm{PL}=\frac{(4\pi)^3r^{4}}{\lambda^2}$, and the resulting \ac{SNR} at the radar receiver is given by 
\begin{equation}\label{eq:SNR-Formula}
\mathrm{SNR}=\frac{\lambda^2\sigma_{\mathrm{rcs}}}{\left(4\pi\right)^3r^4}\frac{\Pav}{\sigma_w^2}\,,
\end{equation}
 where $d$ is the distance between Tx and Rx, the \ac{AWGN} has a \ac{PSD} of $N_{0}$ in W/Hz and other parameters are defined as in \eqref{eq:Chanel_coeff}.



\subsection{Simulation Results}
\begin{figure}[t!]
\centering
\includegraphics[scale=0.85]{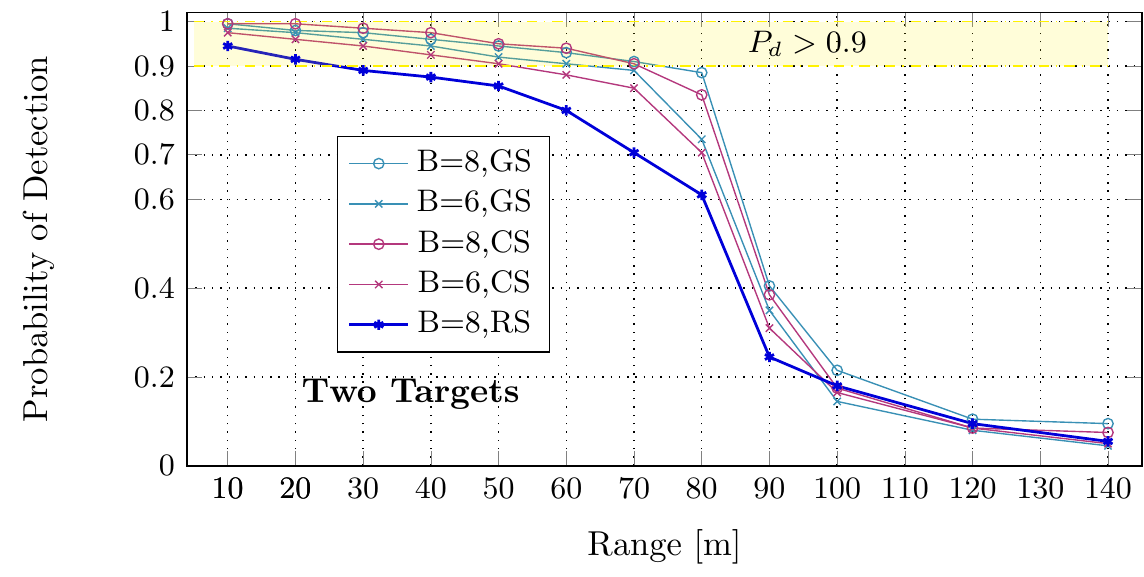}
\caption{Probability of detection vs. range for various number of Blocks for \textbf{G}rid \textbf{S}hift, \textbf{C}ircular \textbf{S}hift, and Random \cite{OTFS_MIMO_AV_22} schemes. Range and SNR can be translated via \eqref{eq:SNR-Formula}.}
\label{fig: Pd_vs_B}
\end{figure}

\begin{figure*}[h]
\centering
\scalebox{.4}{\input{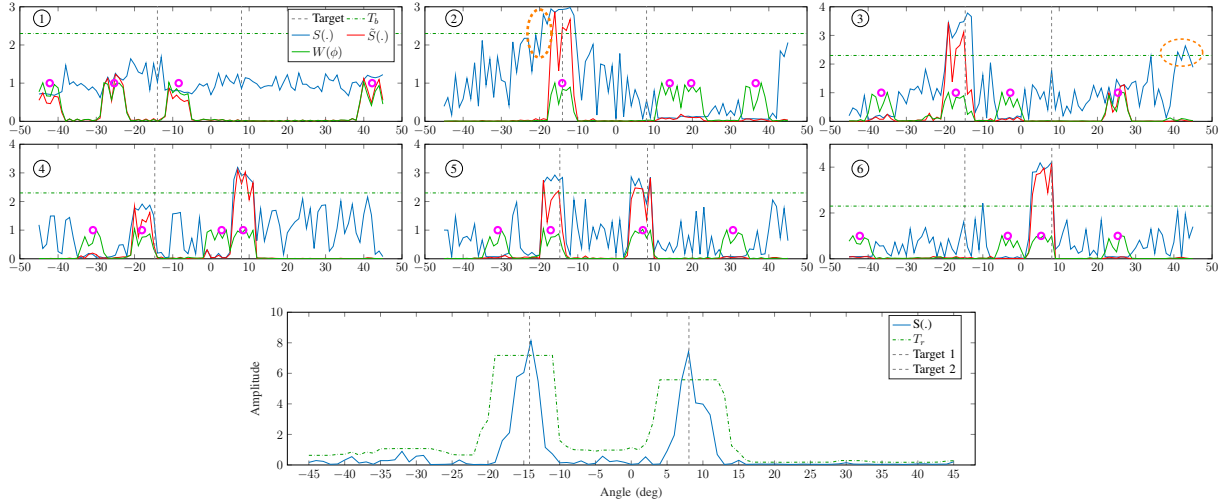}}
\caption{Block-wise depiction of the GS scheme. ML estimates of received signals  $S_{b}(\hat{\nu},\hat{\tau},\phi),~ b\in \{1,..,6\}$ (blue lines) and the weighted version as in \eqref{block-level-prob_BP} (red). The weight function $W(\phi)$ is shown as green curves. At each block $\Nrf=4$ beams are active. The circles show the beam centers used at each block. (Right) The final ML estimate $S(\hat{\nu},\hat{\tau},\phi)$ integrated over $B=6$ blocks with sharpened peaks and OS-CFAR threshold.}
\label{fig: plot_multi_block_Strg1}
\end{figure*}

First, we analyze the detection performance of the proposed active sensing schemes. Fig.~\ref{fig: Pd_vs_B} illustrates the detection probability $P_{\rm d}$ for two targets, placed in a more challenging configuration where they are separated only  in angle domain, as a function of range by varying the number of integration blocks $B$, where $P_{\rm d}=\frac{1}{P}\sum_{p=0}^{P-1}P_{\rm d}(p)$ and $P_{\rm d}(p)$ denotes the detection probability of the $p$-th target. For detection and estimation of parameters, a suitably-defined 3-D grid over $(\nu,\tau,\phi)$ is evaluated (eq. \eqref{log-likelihood2S}) to localize the targets. Since the aim in the \ac{BA} stage is only to obtain an initial estimate of the target parameters, a highly-refined grid is not necessary. Due to the above configuration of targets, Fig.~\ref{fig: plot_multi_block_Strg1} only depicts examples of the received signal at each block, as well as the weighted signal in \eqref{block-level-prob_BP} and the hard detection thresholds, in the angular domain. The \textit{peak-sharpening} effect of the proposed strategy can be observed in the final integrated signal. For the adaptive schemes we have set $L_{\rm max}=3$ (i.e. 3 grid or circular shifts).  A target is considered to be correctly detected if the estimated AoA, $\hat{\phi}_p$ fulfills $\mid \hat{\phi}_p- \phi_p\mid \leq \epsilon$, where $\epsilon$ is set to $0.5^{\circ}$. The curves are obtained by 200 simulation steps at each range, where the angles of targets are randomly changed within an FoV$=[-48^{\circ},48^{\circ}]$. The following observation can be made: 1) The CS method tends to perform better with larger $B$ and at higher \ac{SNR}. 2) The CS method performance improves more drastically with increasing $B$ as compared to the GS technique. 3) When in the low \ac{SNR} regime, all schemes perform very similarly. This behaviour is expected because due to few detections, all schemes are essentially performing random sampling. 
\noindent \subsubsection*{Acquisition Time Comparison}
\begin{figure}[h]
\centering
\includegraphics[scale=0.85]{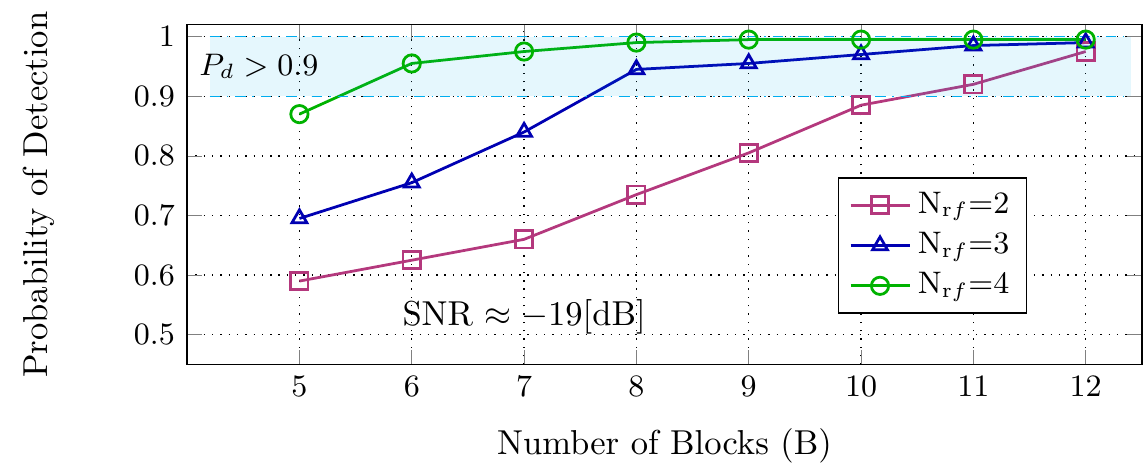}
\caption{Probability of detection vs. number of blocks for GS active scheme. The SNR is a fixed value for the simulation.}
\label{fig: RFChain_vs_B}
\end{figure}

Fig.~\ref{fig: RFChain_vs_B} shows the detection probability $P_d$  as a function of number of blocks $B$ for a single target for a fixed \ac{SNR} value using the GS scheme. Fewer RF chains lead to less sections of the beamspace being scanned at each block and therefore a larger number of blocks are required to achieve the same $P_d$. Note that $B$ can be directly translated to acquisition time $T_a$, by $T_a=NB/\Delta f$.   

\begin{remark}
Analysis of erroneous detections and robustness: When a beam is falsely selected as active, then two shifted variants are selected (e.g. left/right,two circular shifts ). If neither of the two leads to  detections, we conclude that no target exists. The algorithm continues to select other beams in a random manner. In such a case, the two \textit{misspent} beams differ from the original in that they are either grid or circularly shifted versions of the original. Therefore they provide a different projection of the beam space. The misspent beams only lead to a slight increase in acquisition latency.
\end{remark}


\section{Conclusions}\label{sec:Conclusions}
In this work we presented two approaches for active sensing schemes for \ac{HDA} MIMO architectures, especially suitable for \ac{mmWave} \ac{ISAC} applications. Both approaches, prove to be effective in improving the detection probability while reducing the inevitable latency induced by the limited number of available RF Chains. Additionally, these schemes can be adopted to variable number of available RF chains. These approaches can be implemented without a significant increase in codebook complexity and maintain an acceptable \ac{BF} gain over the intended operational range. Our simulations show that even in multi-target scenarios, the schemes are reliably able to detect the targets.

\section{Acknowledgment}
The work of Saeid K. Dehkordi is supported by the German Federal Ministry of Education and Research (BMBF) within the 6G Research and Innovation Cluster 6G-RIC under grant 16KISK030.

\appendices

\bibliography{IEEEabrv,book}



\end{document}